\documentclass[12pt]{iopart}
\usepackage{graphicx}
\begin{document}

\title[Muon Charge Ratio in EAS]{The muon charge ratio in cosmic ray air showers}

\author{H~Rebel$^a$, 
O~Sima$^b$, 
A~Haungs$^a$, 
C~Manailescu$^b$, 
B~Mitrica$^c$, 
C~Morariu$^b$
}
\address{$^a$ Institut f{\"u}r Kernphysik, Forschungszentrum Karls\-ruhe, 
Germany\\
$^b$ Department of Physics, University of Bucharest, Romania\\
$^c$ National Institute of Physics and Nuclear Engineering, Bucharest, Romania}
\ead{rebel@ik.fzk.de}

\begin{abstract}
The muon charge ratio of the lateral muon density distributions in single 
EAS is studied by simulations, in context of recent proposals to measure this observable in coincidence with EAS observations. While effects of the hadronic interaction do not lead to significant differences of the total $\mu^+$ and $\mu^-$ content, the  differences of  the azimuthal variation of the muon densities of opposite charges  and the azimuthal  variation of the muon charge ratio appear to be very much pronounced, dependent on the direction of EAS incidence. This is due to the influence of the geomagnetic field which induces related effects in radio emission from extended air showers.
\end{abstract}

\pacs{96.50.sd, 95.85.Ry}
\vspace{2pc}
{\it Keywords}: cosmic rays, extensive air showers, simulations, muons \\

\vspace{8pc}
\submitto{\JPG}
\maketitle

\section{Introduction}
Primary cosmic rays are dominantly high-energy protons, alpha particles and heavier nuclei  in a relative amount decreasing with the atomic number. When penetrating from the outer space into the Earth's atmosphere they initiate the development of a phenomenon called Extensive Air Showers (EAS) by multiple production of particles in cascading interactions of the primary particles with atmospheric nuclei. The produced secondary radiation establishes an essential feature of our natural environment. It affects material and biological substances and comprises a specific part of the natural radiation background. Its study is of relevance in various scientific problems. Photons, electrons and positrons are the most numerous secondary particles in an EAS event, and the muonic component contributes only to few percent in the single shower. However, at lower primary energies, which are dominating due to the steeply falling primary spectrum, or in case of very inclined showers the electromagnetic shower component gets completely absorbed during the travel through the atmosphere, while the muons (`penetrating component') survive the propagation  through  even larger  slant depths. Hence the inclusive secondary radiation flux in the atmosphere comprises mainly muons (c. 80\% with c.100 muons s$^{-1}$m$^{-2}$sr$^{-1}$ on sea level). \\
The cosmic ray muons originate from decay of hadronic secondaries produced in particle cascades by primary cosmic rays: 
\begin{equation*}
\pi^{\pm} \rightarrow \mu^{\pm} + \nu_\mu (\overline{\nu}_\mu ) ~ ~ ~ 100.0 \% ~ ~ ({\rm mean~lifetime}~ 2.6 \cdot 10^{-2} \mu s) 
\end{equation*}
\begin{equation*}
K^{\pm} \rightarrow \mu^{\pm} + \nu_\mu (\overline{\nu}_\mu ) ~ ~ ~ ~ 63.5 \% ~ ~ ({\rm mean~lifetime}~ 1.2 \cdot 10^{-2} \mu s)
\end{equation*}
\noindent The muons decay with a larger mean lifetime ($2.2 \mu s$):
\begin{equation*}
\mu^+ \rightarrow e^+ + \nu_e + \overline{\nu}_\mu
\end{equation*}
\begin{equation*}
\mu^- \rightarrow e^- + \overline{\nu}_e + \nu_\mu
\end{equation*}
The ratio of the flux of positive to negative muons, the so-called 
muon charge ratio $R_\mu (\mu^+ / \mu^- )$ is a significant quantity which reflects 
important features of the hadronic meson  production in cosmic ray collisions~\cite{ref1,ref2} and can help to discern the primary mass composition~\cite{ref2,ref3}. \\

It is also immediately obvious that the  muon  flux  in  the  
atmosphere is strongly related to the neutrino flux and that the muon 
charge ratio
\begin{equation*}
R_\mu (\mu^+ / \mu^- ) \sim R( \nu_e / \overline{\nu}_e )
\end{equation*}
\noindent provides relevant information for neutrino physics. 
The atmospheric neutrino measurements with 
Super-Kamiokande~\cite{ref4} and other experiments~\cite{ref5} have revealed
that the ratio of muonic to electronic neutrinos is much smaller   
than the theoretical predictions,
$R(\nu_\mu/\nu_e)_{observed} / R(\nu_\mu/\nu_e)_{predicted} \ll 1$. 
The deficit was interpreted in terms of neutrino flavour 
oscillations, confirmed by the observed zenith angular dependence of the measured rates of $\nu_\mu$ and $\nu_e$. \\

There are numerous studies of the charge ratio of the atmospheric muon flux, 
on sea level (see the compilations~\cite{ref2,ref6,ref7,ref8} and also of the vertical dependence by balloon experiments e.g.~\cite{ref9,ref10,ref11}. The experimental results provide a highly inclusive information since the atmospheric flux is produced by many different EAS, with primary energies distributed along the steeply falling energy spectrum and the mass composition of the primary flux. When impinging on our Earth's atmosphere they are additionally affected by the geomagnetic field. The influence of the geomagnetic field leads to a dependence of the muon charge ratio from the azimuth of the direction of observation~\cite{ref12} (East-West effect), in particular for low energy muons, which are dominantly originating from EAS of lower primary energies. \\
The measured value of the muon charge ratio which is empirically found to have a value of about $1.25-1.30$, is mainly a result of the positive charge (proton) excess of the primary mass distribution, at $E_\mu > 10\,$GeV slightly increasing with  the energy of the observed muons~\cite{ref8}. The MINOS detector in the Soudan mine has recently published precise results about the charge ratio of atmospheric muons in the TeV energy range and observed significantly higher values of $R_\mu(\mu^+/\mu^-)$~\cite{ref13}. The rise in the muon charge ratio is expected due to the increasing contribution of kaons to the cosmic ray muon flux and due to an enhanced contribution of the $K^+$ decay~\cite{ref14,ref15}. \\

Modern theoretical approaches of the muon flux and charge ratio start from Monte Carlo simulations of a sufficient number of single EAS events calculated along the primary energy spectrum and chemical distribution of cosmic rays (see ref.~\cite{ref1} e.g.). The Monte Carlo simulations invoke as generators specific models of the hadronic interactions, which are reflected by the so called energy-weighted moments~\cite{ref14}. Therefore measurements of the charge ratio are a source of information about the validity of hadronic interaction models. In case of the atmospheric muons it has turned out that the results are rather stable against modifications of the models, except when higher energy muons would be considered. M.~Unger (L3 Collaboration) has put attention~\cite{ref16} that for muons with 
$E_\mu > \approx 300\,$GeV the charge ratio would be very discriminative, also due to the increasing influence of kaons. The features of the particular models are expected to be displayed  more distinctly, when the charge ratio of single showers is studied. Apparently there is a small  excess of positive pions already in the single collision process, which leads already in a single shower to a value of $R_\mu(\mu^+/\mu^-) > 1$ (see~\cite{ref15}). \\

In exclusive observation of single EAS which could be specified by a definite primary energy, the direction of incidence and eventually by the mass of the primary, the effects of the different types of the hadronic interactions and of the geomagnetic field are expected to get revealed in a more pronounced way. In particular the lateral distribution of the EAS particles displays an azimuthal variation~\cite{ref17}, which is influenced by the geomagnetic field, differently for $\mu^+$ and $\mu^-$, and leading to an azimuthal variation of the charge ratio of the muon density 
distribution. \\

In this paper we address some aspects of the muon charge ratio of single EAS, especially of the azimuthal variations of the muon density distribution, 
worked out by simulation studies.
Few and far between this topic has been addressed in the past~\cite{ref18}.  
The considerations are in context of recent proposals for studies of the muon charge ratio in single EAS, which may be observed with standard detector arrays as used (see ref.~\cite{ref19}) to study EAS. 
The effects depend on the direction of incidence and increase with the inclination of shower incidence (zenith angle $\theta$), i.e. when the distance of muon travel is increased due to the $sec(\theta)$-dependence. Recently, in view of the possibility to extract charge information of high energy EAS muons the geomagnetic effects on the shower development have been estimated in~\cite{ref20,ref21} on basis of a modified  Heitler model~\cite{ref22}, originally known as `toy-model'. 
The present study yields results on the lateral muon distributions resulting from extensive three-dimensional Monte Carlo simulations with reconstructions of the azimuthal variation of the muon component of inclined showers (with $\theta=45^\circ$ as example and with the primary energies of $10^{14}\,$eV and $10^{15}\,$eV).

\section{Azimuthal effects of lateral distributions of EAS charged particles}

\begin{figure}
\begin{center}
\includegraphics[width=14.cm]{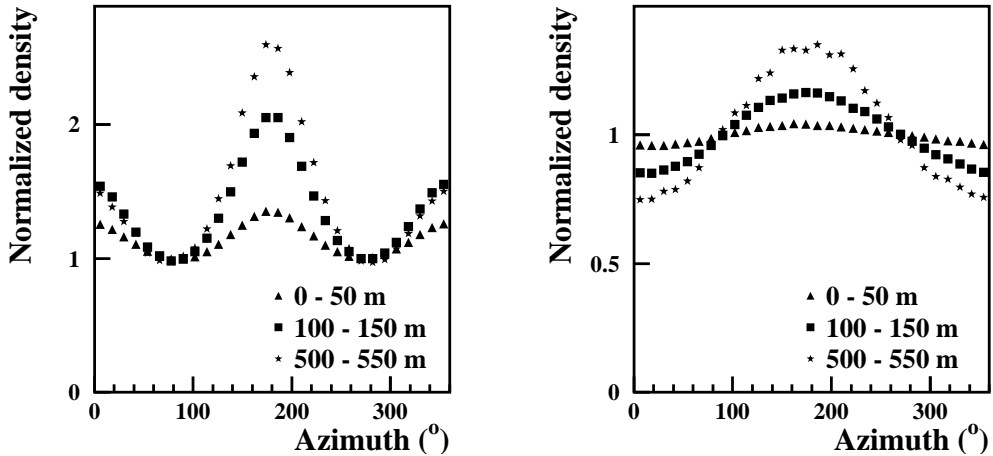}
\end{center}
\caption{\it The azimuthal distribution of the normalized charged particle 
density of high energy proton induced EAS ($E=6.85 \cdot 10^{17}\,$eV, $\theta=37^\circ$, $\phi=40.3^\circ$) in the observation plane (left) and in the plane normal to the shower axis (right). The normalization factor is the average value of the density for the azimuth $\phi=90^\circ$ and $\phi=270^\circ$, ranging from approx. 8500 particles/m$^2$  (radial range 0-50$\,$m) to 85 particles/m$^2$ (radial range 100-150$\,$m) to 1.24 particles/m$^2$ (radial range 500-550$\,$m) in the observational plane (see text).}
\label{fig1}
\end{figure}

Azimuthal asymmetries and the azimuthal variation of the lateral distributions of charged EAS particles originate dominantly from the attenuation and geometrical effects of showers with inclined incidence. For a qualitative explanation we may approximate that the EAS start from infinity and neglect any influence of the geomagnetic field in a first step of the argumentation. 
Thus we have cylindrical symmetry around the shower axis for all radial distances 
from the axis. For inclined showers hitting the observation plane, charged particles arriving first (`early' azimuthal EAS region) do experience less attenuation than the particles arriving later (`late' azimuthal EAS region), as the latter have traveled larger distances in atmosphere. 
The azimuthal asymmetries are also present when transforming the observed particle densities to the more relevant normal shower plane, orthogonal to the axis. Such azimuthal asymmetries in the normal plane arise from geometrical effects of the projection procedure and must be corrected in the analyses. 
Recently~\cite{ref17}, in context of investigations of the charge particle EAS component, registered by the measurements of the KASCADE-Grande experiment~\cite{ref23}, the azimuthal asymmetries have been considered in detail 
for EAS of inclined incidence. An example is displayed in Fig.~\ref{fig1}. It shows for a particular case of the direction of EAS incidence the azimuthal variation of the charge particle 
density for various (radial) distances from the shower centre in the plane 
normal to the shower axis and in the observation plane. For sake of convenient comparison the distributions for different radial distances have been normalized to the mean value of the charge particle densities at $\phi=90^\circ$ and  $\phi=270^\circ$. These directions, corresponding to the intersection of the plane normal to the shower axis with observation plane, were chosen for the normalization because the attenuation effects are identical in both directions and also in the observational and normal plane. Here the (intrinsic) azimuth angles $\phi$  are counted counter-clockwise from the direction of the intersection of the vertical 
plane containing the shower axis with the observation plane and the normal plane, respectively ($\phi=0$ corresponds to the late region of the shower). \\

The azimuthal dependence in the normal plane reflects mainly the effect of attenuation, while in the observation plane geometric effects are superimposed 
(points located in the observation plane at the same distance $R$ from the shower core correspond to points located in the normal plane at distances from the core changing from $R \cdot \cos{\theta}$ to $R$ when $\phi$ changes from $0$ to $90^\circ$).

\section{Regarding EAS muons}

In order to explore the azimuthal asymmetries of the lateral distributions, in particular of the muon component, some realistic and detailed Monte Carlo simulations of the EAS development have been performed. For the simulations the program 
CORSIKA~\cite{ref24} (version 6.5) has been used, invoking different models of the hadronic interaction, in particular QGSJET~\cite{ref25}. The Earth magnetic field with a homogeneous field approximation, observation level ($110\,$m a.s.l.), and the particle energy thresholds have been chosen according to the conditions of the KASCADE-Grande experiment ($E_\mu^{\rm thres} = 100\,$MeV). The U.S. standard atmosphere (see ref~\cite{ref24}) has been adopted. For a realistic description the electron-photon component is simulated by the EGS option~\cite{ref26}. \\
The performed simulations comprise proton and Fe induced EAS, with a zenith angle of incidence of $\theta = 45^\circ$ and arriving from different cardinal points: North, East, South, West. About 1000 showers have been simulated for most cases. For each event the muon lateral distributions for both muon charges separately and the distribution of the charge ratio of the muon density have been reconstructed. In order to explore the influence of the geomagnetic field, the simulations have been compared with those when the Earth's magnetic field is switched off. \\

The following figures (Figs.~\ref{fig2} - \ref{fig8}) display a selection of results, mainly for the primary energy of $10^{15}\,$eV, since the EAS of that energy have a muon content large enough leading to a sufficient statistical accuracy. Only the case of protons is displayed as primary Fe show similar features. The distributions are displayed only in the observation plane as function of the azimuth angle $\phi$ around the shower core (with a changed definition of the azimuth angle different from the intrinsic coordinates (see~\cite{ref17}) used in Fig.~\ref{fig2} with the convention used further on: a point located in the North from the shower center has azimuth $\phi = 0$, a point located in East has $\phi = 90^\circ$). Further cases are illustrated in the appendix of reference~\cite{ref15}. \\
\begin{figure}
\begin{center}
\includegraphics[width=7.cm]{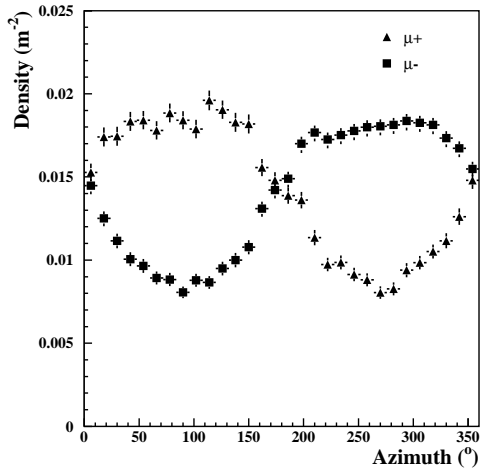}
\end{center}
\caption{\it The mean azimuthal $\mu^+$ and $\mu^-$ distributions of proton induced EAS of inclined showers ($\theta = 45^\circ$) incident from  North with the primary energy of $10^{15}\,$eV at a distance of $45-50\,$m from the shower axis.}
\label{fig2}
\end{figure}
\begin{figure}
\begin{center}
\includegraphics[width=7.cm]{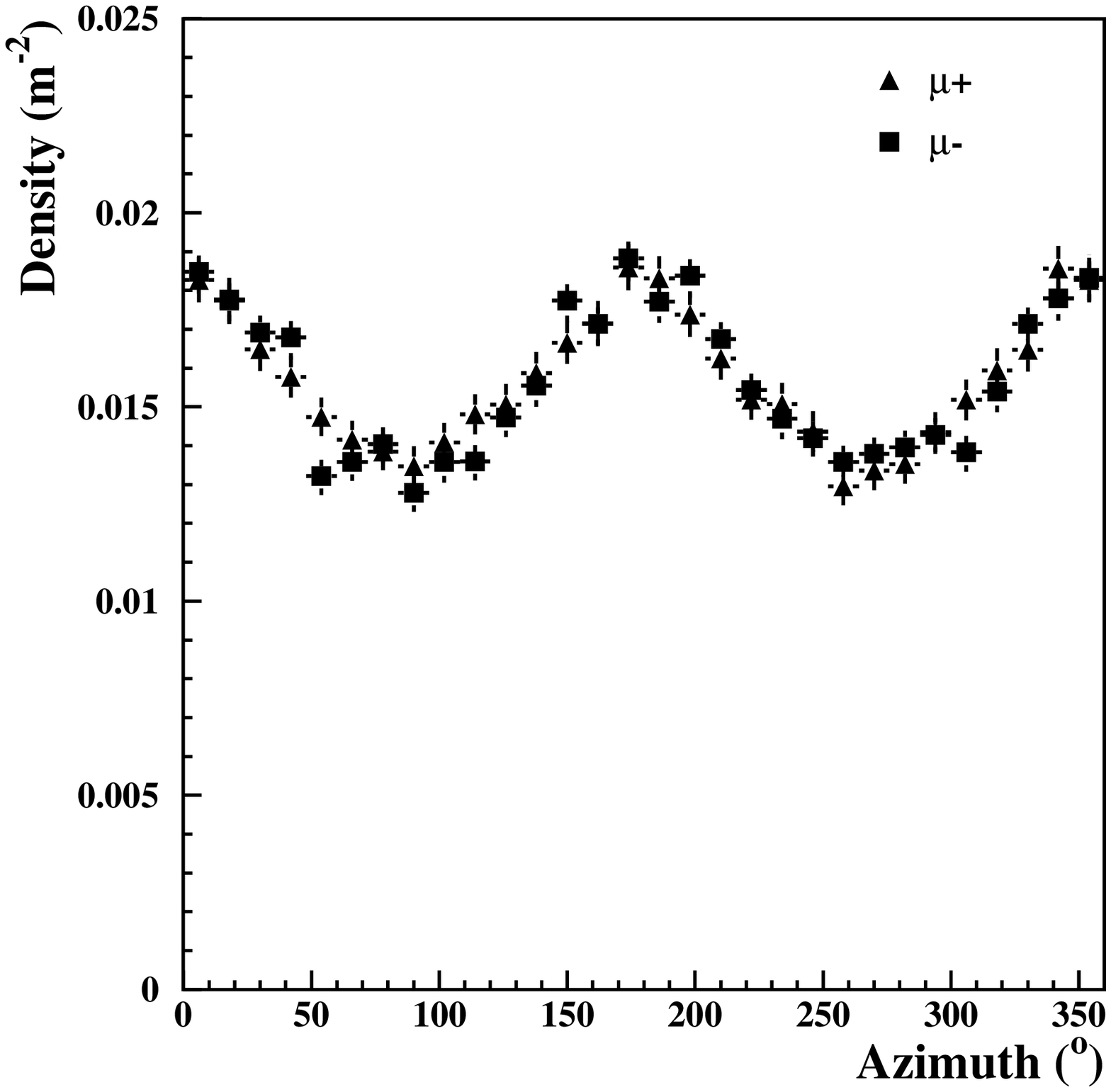}
\end{center}
\caption{\it The mean azimuthal $\mu^+$ and $\mu^-$ distributions of proton induced EAS of inclined showers ($\theta = 45^\circ$) incident from  North with the primary energy of $10^{15}\,$eV at a distance of $45-50\,$m from the shower axis. Earth's magnetic field switched off.}
\label{fig3}
\end{figure}
\begin{figure}
\begin{center}
\includegraphics[width=7.cm]{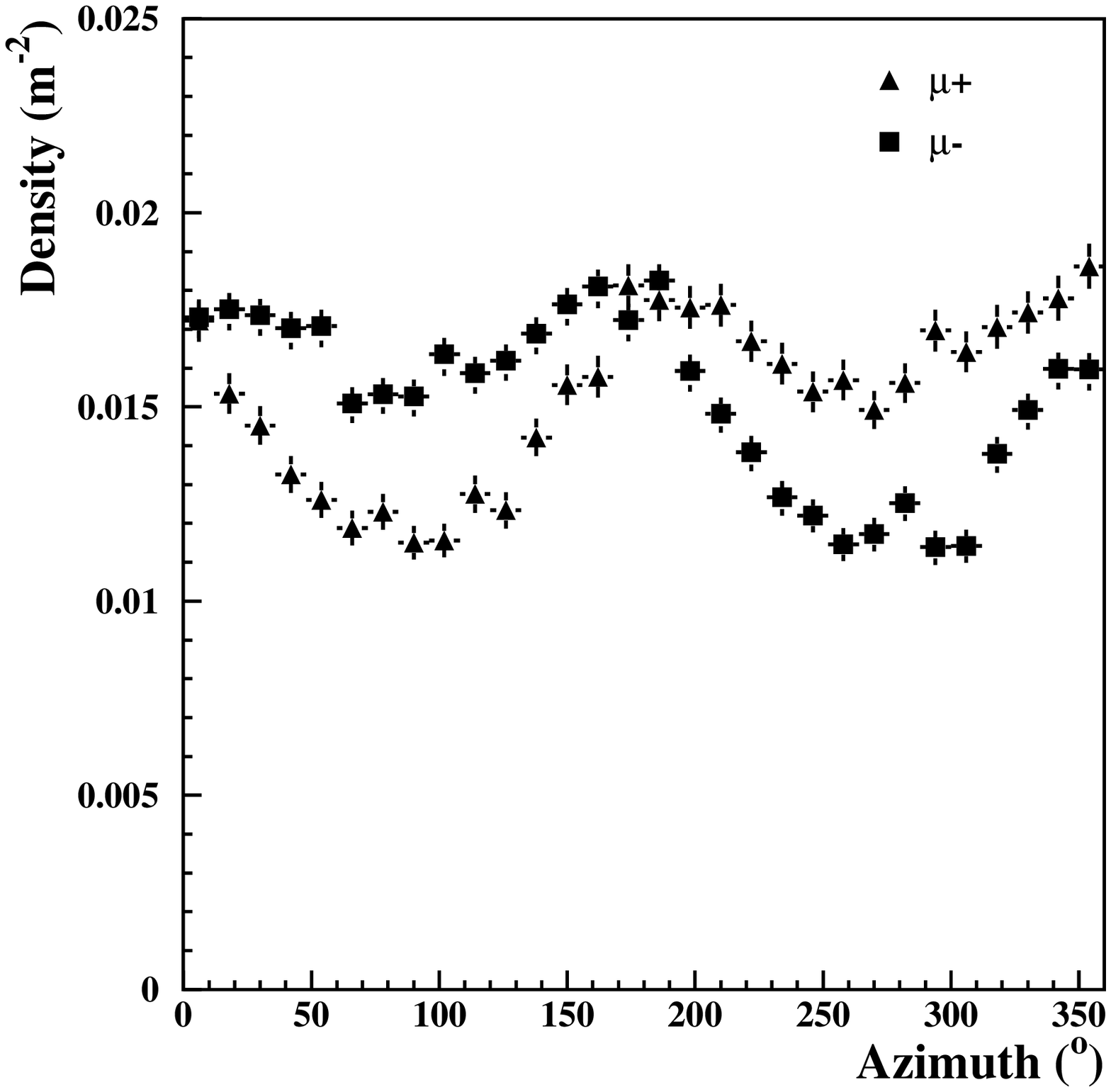}
\end{center}
\caption{\it The mean azimuthal $\mu^+$ and $\mu^-$ distributions of proton induced EAS of inclined showers ($\theta = 45^\circ$) incident from  South with the primary energy of $10^{15}\,$eV at a distance of $45-50\,$m from the shower axis.}
\label{fig4}
\end{figure}
\begin{figure}
\begin{center}
\includegraphics[width=7.cm]{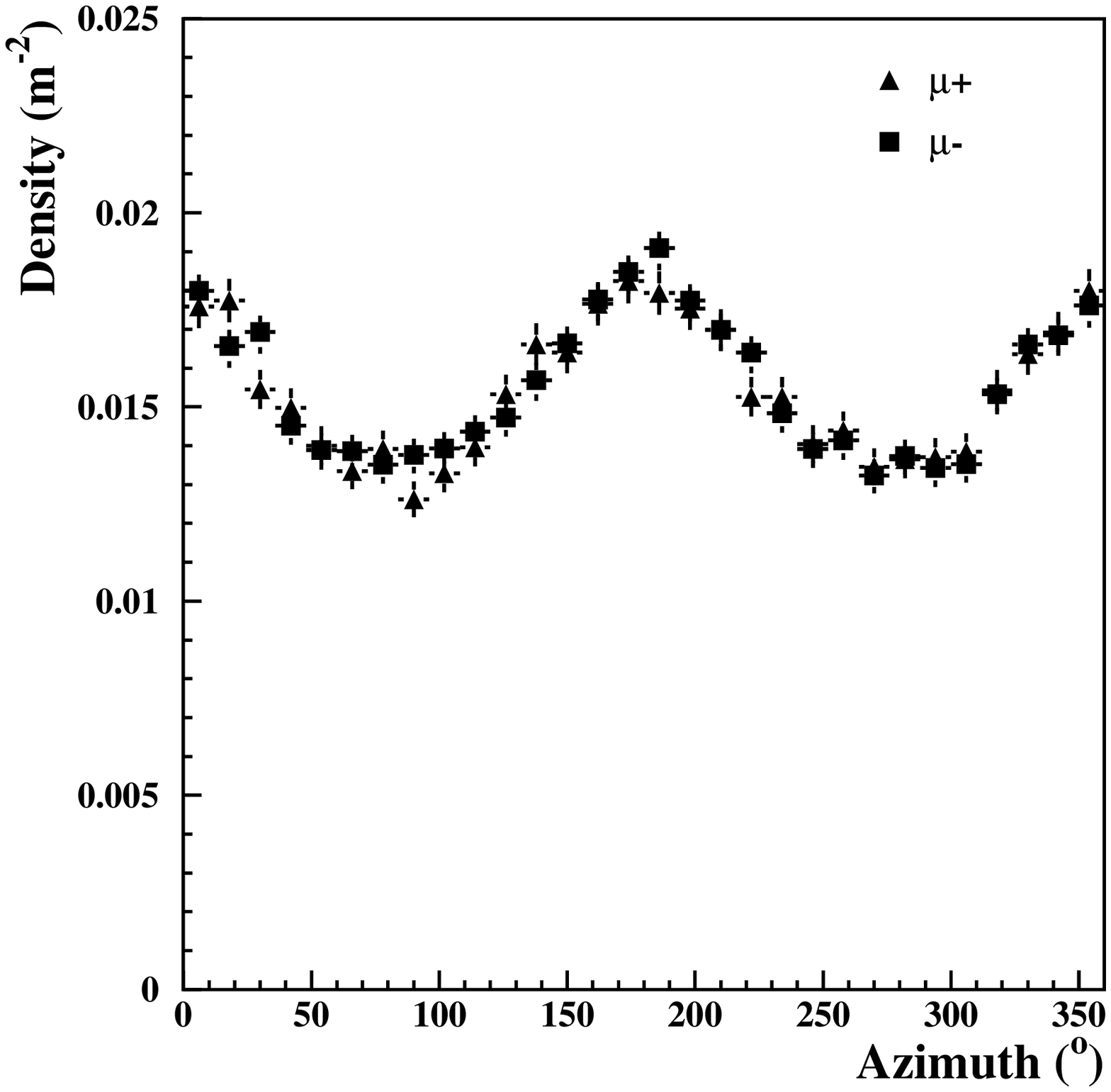}
\end{center}
\caption{\it The mean azimuthal $\mu^+$ and $\mu^-$ distributions of proton induced EAS of inclined showers ($\theta = 45^\circ$) incident from  South with the primary energy of $10^{15}\,$eV at a distance of $45-50\,$m from the shower axis. Earth's magnetic field switched off.}
\label{fig5}
\end{figure}
\begin{figure}
\begin{center}
\includegraphics[width=7.cm]{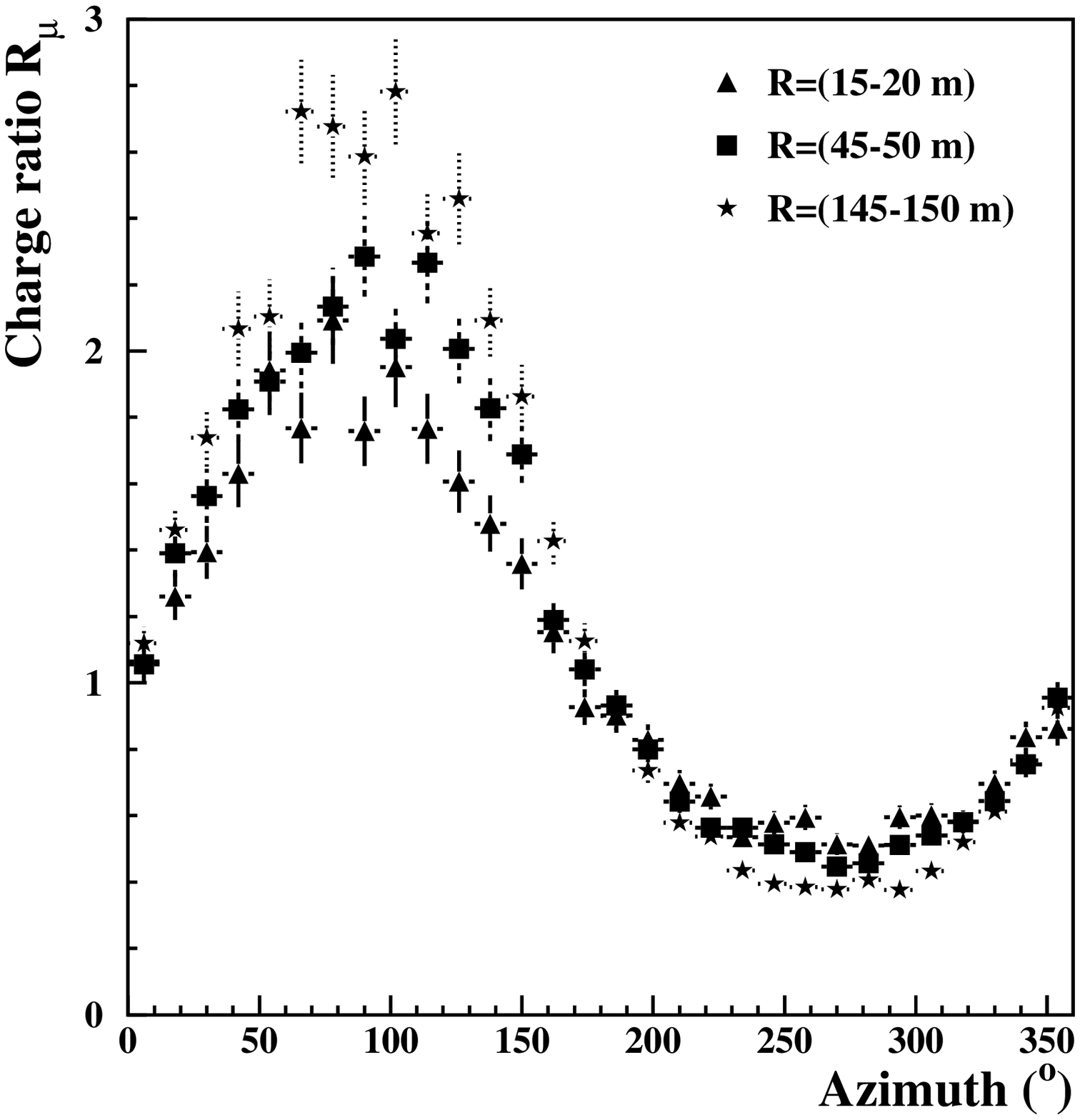}
\end{center}
\caption{\it Variation of the charge ratio $R_\mu$ of the mean muon density distribution of proton induced EAS of inclined showers ($\theta = 45^\circ$) 
incident from North with the primary energy of $10^{15}\,$eV at various 
distances $R$ from the shower axis.}
\label{fig6}
\end{figure}

The features shown in Figs.~\ref{fig2} - \ref{fig8} can be understood when taking into account the direction of the geomagnetic field vector (for the location of Karlsruhe for which the calculations are made, the magnetic inclination has a value of 
c. $\gamma = 65^\circ\,$North, i.e. the vector points downwards). 
Fig.~\ref{fig1} shows the azimuthal variation of the total charge particle component of proton induced showers of higher primary energy. Though the effect of the geomagnetic field is included, it is obscured since positive and negative particles behave in an opposite way, and since in addition  the effect for electron and positrons appears to be less pronounced. The electrons and positrons considerably suffer from scatterings, changing the directions of the momenta relative to the geomagnetic field. \\
Hence the total charge particle distribution resembles the $\mu^+$ and $\mu^-$ distributions, when the geomagnetic field is just 'switched off' (Figs.~\ref{fig3} and~\ref{fig5}). 
In the realistic case, however the influence of the geomagnetic field on the muon distribution is generally pronounced (Fig.~\ref{fig2}). \\
In Fig.~\ref{fig2} the particles are coming from the North. Thus the Lorentz force acting on the path ways of $\mu^+$ is directed towards East and the $\mu^+$-density is enhanced  around $\phi=90^\circ$ (East). 
In case the particles are coming from South, the Lorentz force acts on $\mu^+$
in the opposite direction, shifting the maximum of the distribution to West (Fig.~\ref{fig4}). 
However, quantitavely the effect of the Earth magnetic field is different for muons coming from North and South, respectively, for EAS of the same kind. This feature depends on the magnetic inclination of the geomagnetic field and on the direction of the shower incidence. Thus the relative angle (geomagnetic angle) between the particle momenta ($\theta = 45^\circ$) and the magnetic vector amounts to be about $70^\circ$ for particles arriving from North, while for particles arriving from the South the value of the geomagnetic angle is $20^\circ$. This situation explains the different amplitudes of the azimuthal variation of the charge ratio in Figs.~\ref{fig6} and~\ref{fig7}. \\
\begin{figure}
\begin{center}
\includegraphics[width=7.cm]{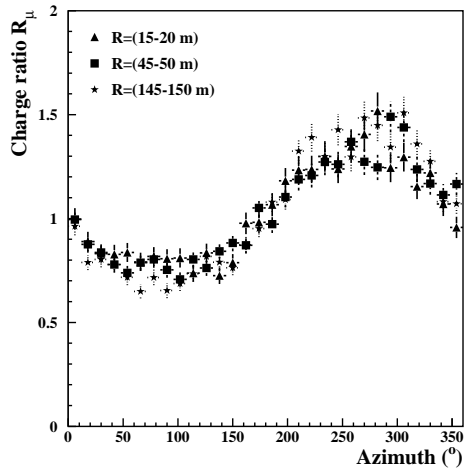}
\end{center}
\caption{\it Variation of the charge ratio $R_\mu$ of the mean muon density distribution of proton induced EAS of inclined showers ($\theta = 45^\circ$) 
incident from South with the primary energy of $10^{15}\,$eV at various 
distances $R$ from the shower axis.}
\label{fig7}
\end{figure}
\begin{figure}
\begin{center}
\includegraphics[width=7.cm]{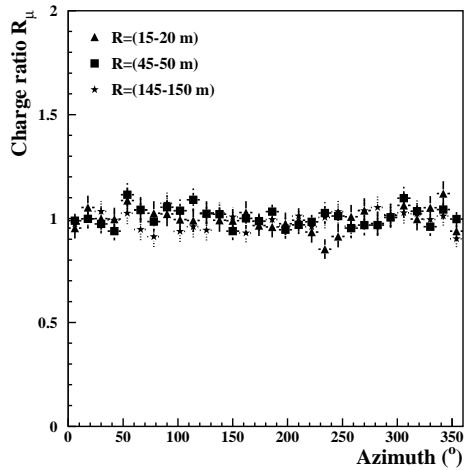}
\end{center}
\caption{\it Variation of the charge ratio $R_\mu$ of the mean muon density distribution of proton induced EAS of inclined showers ($\theta = 45^\circ$) 
incident from North with the primary energy of $10^{15}\,$eV at various 
distances $R$ from the shower axis. Earth's magnetic field switched off.}
\label{fig8}
\end{figure}

The azimuthal $\mu^+$ and $\mu^-$ distributions, calculated with the geomagnetic field `switched off' and displayed in the observation plane, shown in Figs.~\ref{fig3} and~\ref{fig5}, are  mainly the results of geometric effects. The density at $\phi = 0$ and $\phi = 180^\circ$ is larger than at $\phi = 90^\circ$ and $\phi = 270^\circ$ since for the density at the radial distance $R$ from the shower center in the observation  plane corresponds to the density at the radial distance $R \cdot \cos{\theta}$ in the normal plane at $\phi = 0$ and $\phi = 180^\circ$, but that at $R$ for the densities at $\phi = 90^\circ$ and $\phi = 270^\circ$. The observation that the density at $\phi = 0$ and $\phi = 180^\circ$ are practically identical indicates that the attenuation of the particle density is of minor importance for the muon distribution in the distance
$R= 45-50\,$m. At larger distances also attenuation effects get evident. \\
When increasing the  distance from the shower core, the amplitude of the azimuthal variation of the charge ratio (Figs.~\ref{fig6} and~\ref{fig7}) is increasing. 
This is due to the distortion of the lateral distribution, noticeably with increasing radial distance. For EAS  arriving from the North the  $\mu^+$ density get shifted 
to larger distances from the shower center in East direction, i.e the density corresponds to the density of smaller radii of the undisturbed lateral distribution. In case of $\mu^-$ density corresponds there to the undisturbed lateral distribution of larger distances. \\
For EAS arriving from South this happens in West direction. There are experimental observations~\cite{ref27} of the KASCADE-Grande experiment~\cite{ref23} which indicate similar features. \\

From these features and from tentative results of studies at lower primary 
energies we deduce the following conclusions:
\begin{enumerate}
\item	The azimuthal variation of the muon density and of the local muon charge 
ratio in EAS are strongly influenced at the considered energies by the 
geomagnetic field (see Figs.~\ref{fig2},~\ref{fig4}, and~\ref{fig6}).
\item	This influence depends from the azimuthal direction of the EAS incidence,
i.e the direction relative to the direction of the Earth magnetic field and 
it increases with the distance from the shower axis (compare Fig.~\ref{fig6} and Fig.~\ref{fig7}).
\item In case of neglecting the geomagnetic field the azimuthal variation in 
the observation plane reduces to the variation due to geometry effects and for
larger radial distances from the core also due to the attenuation effect (see Fig.~\ref{fig1}) which does not influence differently positive and negative muons 
(consequently the integrated value of $R_\mu$ remains nearby unity, since obviously 
differences in the $\pi^+$ and $\pi^-$ production remain rather small at the 
considered muon energies; see Fig.~\ref{fig8}).
\end{enumerate}

\begin{figure}
\begin{center}
\includegraphics[width=14.4cm]{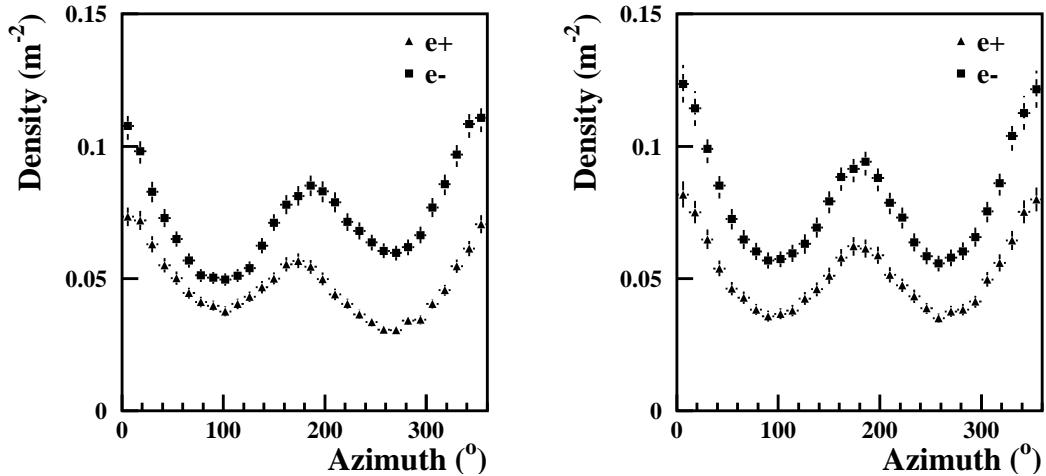}
\end{center}
\caption{\it The mean azimuthal $e^+$ and $e^-$ density distributions of proton induced inclined EAS ($\theta = 45^\circ$) incident from North with the primary energy of $10^{15}\,$eV at a distance of $45-50\,$m from the shower axis. The case when the geomagnetic field is ignored, is shown on the right panel.}
\label{fig9}
\end{figure}

Fig.~\ref{fig9} displays for a particular case the effect of the geomagnetic field on the electron-positron components. Though rather reduced and dominated by attenuation, geomagnetic effects are not completely obscured. \\

It should be additionally remarked that for proton induced showers (incident from North or South) the present studies result in the muon charge ratio integrated over all 
distances to $R_\mu =1.028 \pm 0.002$ (for Fe induced EAS in 
$R_\mu = 1.025 \pm 0.0015$), where the uncertainty is the standard error for a typical EAS. 

\section{Influence of the EAS muon energy and of the zenith angle of EAS incidence}

\begin{figure}
\begin{center}
\includegraphics[width=14.cm]{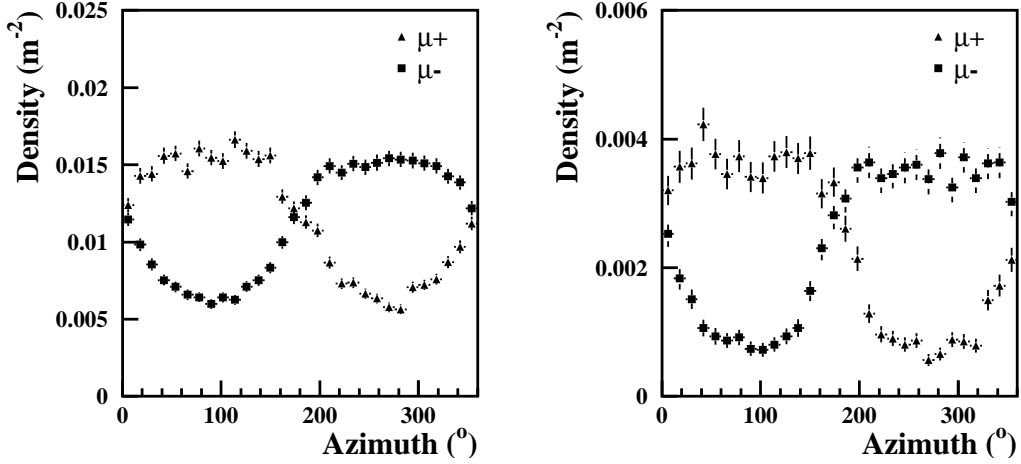}
\end{center}
\caption{\it The mean azimuthal $\mu^+$ and $\mu^-$ distributions of proton induced EAS of inclined showers ($\theta = 45^\circ$) incident from 
North with the primary energy of $10^{15}\,$eV at a distance of 
$45-50\,$m from the shower axis: Comparison of different muon energy 
thresholds: $10\,$GeV (left panel) and $100\,$GeV (right panel).}
\label{fig10}
\end{figure}

\begin{figure}
\begin{center}
\includegraphics[width=14.cm]{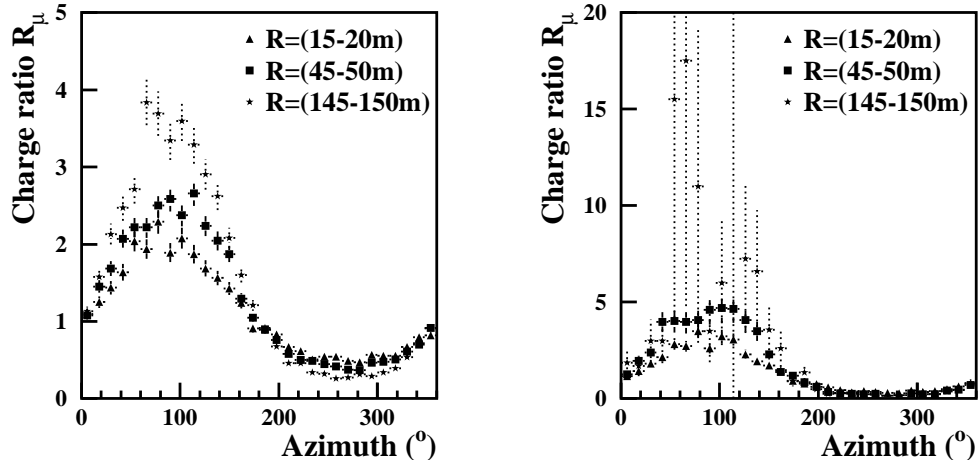}
\end{center}
\caption{\it Variation of the charge ratio $R_\mu$ of the mean muon density of proton induced EAS of inclined showers ($\theta = 45^\circ$) incident from 
North with the primary energy of $10^{15}\,$eV at different distances from the shower axis: Comparison of different muon energy thresholds: $10\,$GeV (left panel) and $100\,$GeV (right panel).}
\label{fig11}
\end{figure}

The energy spectrum of EAS muons has a rather flat maximum at energies of some few GeV, before decreasing, say from $2-3\,$GeV on. Consequently the features shown for a muon energy threshold  of $0.1\,$GeV do not change very much when increasing the threshold to $1\,$GeV, at least not for the considered central range of radial distances from the shower center, since the fraction of muons below $1\,$GeV is only in the order of 10\%. The features appearing when muons with energies larger than $10\,$GeV and $100\,$GeV, respectively, are considered, are shown in Figs.~\ref{fig10} and~\ref{fig11} for the case of EAS arriving from the North.
The graphs which can be compared with Figs.~\ref{fig2} and~\ref{fig6} indicate also the strong decrease of the muon density with increasing the muon energy beyond few GeV. This 
very low density of TeV muons in individual EAS
does cast some doubt on the practicability of attempts to use the Earth magnetic field as charge separator of very high energy muons from very inclined EAS~\cite{ref20,ref21}. The variation of the muon charge ratio, though globally with the same features like at lower energies, displays increasing amplitudes with increasing the muon energy. \\

\begin{figure}[ht]
\begin{center}
\includegraphics[width=7.cm]{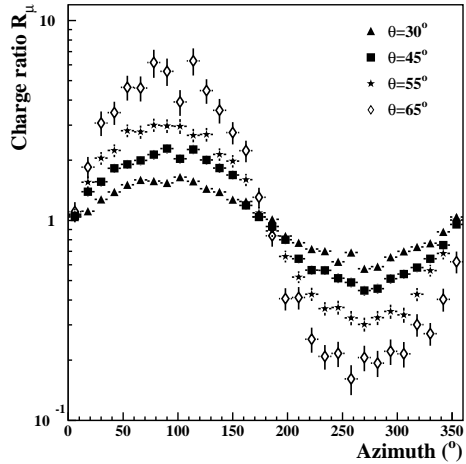}
\end{center}
\caption{\it Azimuthal variation of the charge ratio $R_\mu$ of the muon density of proton induced EAS incident with different zenith angles from North with the primary energy of $10^{15}\,$eV, observed at a radial distance of $45-50\,$m and for a muon energy threshold of $0.1\,$GeV.}
\label{fig12}
\end{figure}

The influence of the geomagnetic field and the separation of $\mu^+$ and $\mu^-$ increase with the path length (slant depth) of the muon trajectories in the atmosphere. Hence the azimuthal $R_\mu$ variation gets more pronounced with increasing distances from the shower core, with the threshold of observed muon energies since muons of higher energies stem from earlier generations, and with the zenith angle of EAS incidence. This is displayed in Fig.~\ref{fig12}.

\section{Remarks on similar features of radio emission from EAS}

The electromagnetic component of extensive air showers is accompanied  by various types of secondary radiation arising from different interaction mechanisms. Recently the emission of radiation in the radio frequency range (`radio flashes') from EAS has been re-discovered~\cite{ref28}, which most likely originates from the so called coherent geosynchrotron effect: electron positron pairs get bent in the terrestrial magnetic field and are emitting synchrotron radiation. This discovery, already indicated by early observations in the sixties, recently definitively confirmed by the advanced detection techniques, has prompted considerable activities, e.g. the installation of LOPES~\cite{ref29}. That is an array of a number of dipole antennas placed inside the particle detector array of KASCADE-Grande for the detection of radio emission in coincidence with air showers observed with particle detectors. According to analytical studies~\cite{ref30} a broadband short radio pulse in the nanosecond range is expected, observable on ground with $5-15\,$V/m/MHz for EAS of primary energy of approximately $10^{17}\,$eV. In fact such radio pulses have been observed with the digital radio antenna field of thirty East-West polarised short antennas located within the KASCADE-Grande field. \\
\begin{figure}
\begin{center}
\includegraphics[width=9.4cm]{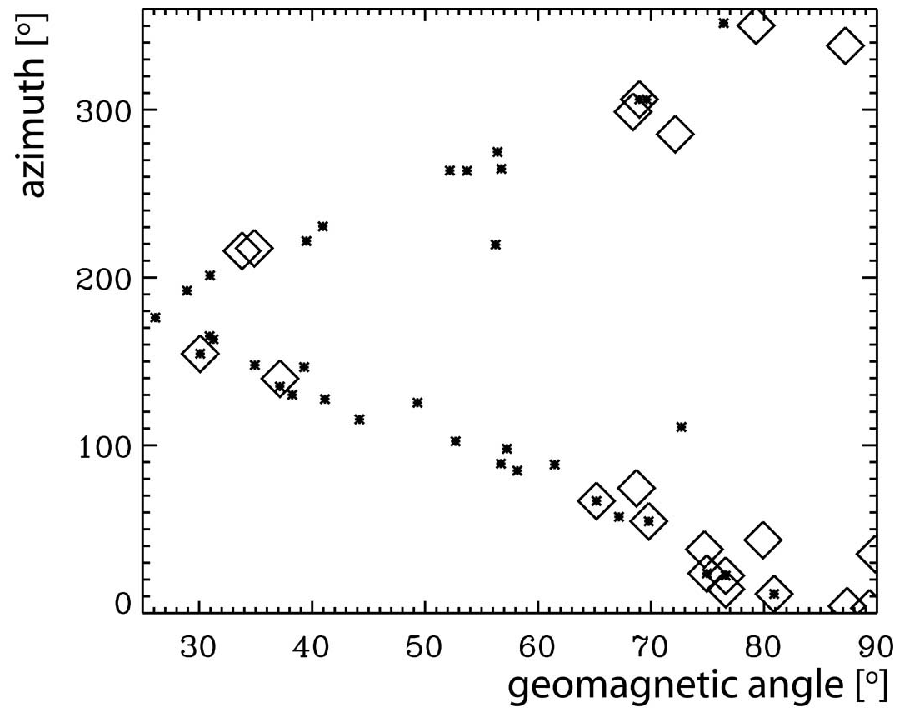}
\end{center}
\caption{\it Measured inclined air showers as a function of azimuth and geomagnetic angles of the incident EAS
(from ref.~\cite{ref33}). Crosses represent events which are reconstructed from KASCADE-Grande observations, rhombs indicate events detected by radio signals. The azimuth convention is the same as in  previous graphs: $90^\circ$ and $270^\circ$ azimuth angles denote showers coming from East and West, respectively, $180^\circ$ and $0/360^\circ$ denote showers arriving from South and North, respectively.}
\label{fig13}
\end{figure}

The assumed geomagnetic origin of the expected emission process suggests a dependence of the resulting field strength at observation position on the direction of the terrestrial magnetic field like the effects of the geomagnetic field on the variation of the EAS muon density. Indeed, a strong correlation of the field strength with the angle between Earth magnetic field and shower axis (geomagnetic angle) is expected~\cite{ref31}. But following simulation studies, the dependence on the strength and geometry of the magnetic field appears to be more subtle than intuitively expected, and detailed effects show up in the polarization characteristics of the radio signal. Simulations~\cite{ref32} let expect that the emission and the radio signal strength - like the EAS muon density - depend as well on the zenith as on the azimuth angles and not simply only on the geomagnetic angle. Thus inclined showers with large zenith angles exhibit a much larger radio footprint at observation level, making them especially favorable for radio detection. This feature is analogous to the separability of positive and negative muons by the Earth's magnetic field with increasing zenith angle. The azimuthal angle, which is of course related with geomagnetic angle at a given observation location and zenith angle of EAS incidence, defines the degree of the polarization of the measurable signal. In addition to the zenithal shower direction also the (azimuthal) incidence direction of the shower and the position of the observer influences the field strength and the degree of polarization~\cite{ref33}. \\
It is of interest to compare the azimuthal dependence of the radio signal with the  dependence of the muon charge ratio from the azimuthal incidence.  The muon charge ratio provides a measure of  the separation of positive and negative charges by the geomagnetic field. The present experimental status of information from LOPES is displayed in Fig.~\ref{fig13}, where a sample of observed showers with zenith angle of incidence larger than $50^\circ$ is shown. Large zenith angles are considered, not only because of the enhanced appearance of radio signals, but also since a larger range of the geomagnetic angle values could be covered. The figure (taken from~\cite{ref33}) shows KASCADE-Grande triggered and radio detected events in a plot of azimuth angle of incidence vs. the geomagnetic angle. Two features can be noticed.
\begin{itemize}
\item A pronounced North-South asymmetry. There appear more radio detected events  arriving  from the North than from the South, which is probably related to the fact that cosmic rays arriving from North have the largest geomagnetic angles. This feature is analogous to the different amplitudes of the azimuthal variation of the muon charge ratio (and also electron charge ratio) for EAS arriving from North and from South (see Figs.~\ref{fig6} and~\ref{fig7}).
These figures show that the amplitude of the charge ratio variation is more pronounced for showers coming from North than for showers coming from South. For showers coming from North the charge ratio in the Eastern direction from the core is about 2 (Fig.~\ref{fig6}, azimuth 90$^\circ$), while for showers coming from South it is less than 1.5 in the Western direction from the core (Fig.~\ref{fig7}, azimuth 270$^\circ$). The geomagnetic field has an enhanced effect for showers coming from North both in the case of charge ratio amplitudes and in the case of detected radio events.
\item A pronounced gap with no detected radio events (missing 'rhombs' region in 
Fig.~\ref{fig13}). There appear no radio signals from showers coming from East or West ($90^\circ$ and $270^\circ$). This may be explained by the fact that the particular antenna arrangement of LOPES used at that time did only measure the East-West linear polarisation of the radiation. Simulations predict that the radio waves from showers arriving from East or West should be mostly North-South polarised. This effect of the polarisation has of course no analogue in the variation of the muon density.
\end{itemize}
Obviously the influence of the Earth's magnetic field plays in both cases a dominant role for detailed features of the $\mu^+$ and $\mu^-$ density distributions and of the emission of the radio waves from EAS.

\section{Concept of experimental approaches}

Experimental studies of the azimuthal and radial variations of the charge ratio of the muon density in EAS would need a device (spectrometer) for the determination of $R_\mu$ placed inside or nearby a detector array for determining the incident EAS: the direction of incidence, defining also the azimuth position (relative to the plane of shower incidence) of the spectrometer and the position of the shower core relative to the location of the spectrometer. \\
The group of the Okayama University, Japan~\cite{ref34,ref35} has installed on the rooftop of a campus building a small array of 8 plastic scintillators (OU array) distributed over an area of about $15\,$m$\cdot25\,$m in order to observe EAS. In addition there is a magnetic spectrometer (Okayama telescope) located under the OU array to spectroscopy positive and negative muons, especially to determine the charge sign from the curvature of the tracks of the incident muons and the charge excess of the muons with a momentum larger than $1\,$GeV/c. The apparatus could be used for a coincident observation of the EAS array with the magnetic spectrometer. First results of the analyses of measured data indicate a  charge excess, disappearing within the (large) error bars. But there has been neither the azimuth location of the spectrometer relative to the shower plane nor the angles of the incident EAS specified, and the results may be understood as an average cancelling out the effects. Also the fact that the Okayama telescope finds a rather small value of the muon charge ratio of atmospheric muons may cast some doubts on the systematic accuracy of the measurements. \\
\begin{figure}
\begin{center}
\includegraphics[width=8.4cm]{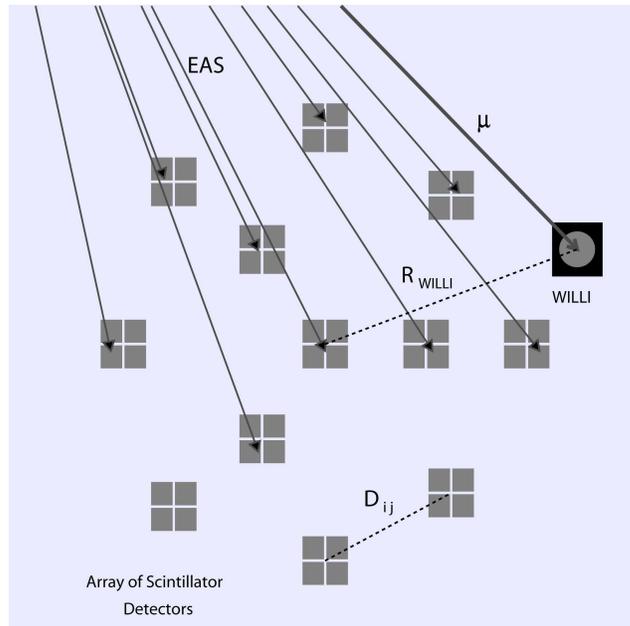}
\end{center}
\caption{\it Sketch of a possible geometrical layout of a mini-array for triggering the 
nearby located WILLI device for muon detection~\cite{ref37}.}
\label{fig14}
\end{figure}

The spectrometer WILLI~\cite{ref2,ref36,ref37} installed in Bucharest, Romania determines the muon charge ratio by measuring the mean life time of the muons stopped in a stack of 16 detector layers. The life time is different for positive and negative muons as negative muons can be captured in atomic orbits, thus leading to an effectively reduced lifetime, depending on the stopping material. The uncertainties of the efficiencies for the detection of differently charged muons and of the geometrical acceptance, usually affecting magnetic spectrometers, are cancelled out. The detector area of the device is able to be directed in a particular direction to measure the incoming muons under different (zenith and azimuth) angles of incidence. Actually WILLI has provided results of successful measurements, resulting in $R_\mu$-values well in the range of the world average, down to a momentum range of about $0.2\,$MeV/c. It has detected the azimuthal variation (East-West effect) of the muon charge ratio of atmospheric muons. \\

There is a proposal (see ref.~\cite{ref37}) to equip the WILLI device with an array of 12 scintillator units of $\approx 1\,$m$^2$ size each, located nearby WILLI, in order to determine the EAS core position relative to the WILLI device and the direction of incidence of showers of  the energy  region  of $10^{14}-10^{15}\,$eV, providing a trigger for WILLI.  
Fig.~\ref{fig14} displays a sketch of a geometrical layout of the detector system which is under consideration by simulation studies with respect to the efficiency and the detector responses~\cite{ref38} (specifying the geometrical layout: $R_{\rm WILLI}$, $D_{ij}$). 
Though such a small array will not be able to determine the energy of the incident showers, it is expected that such detector system will provide more detailed information about the different propagation of positive and negative EAS muons, in particular under the influence of the geomagnetic field.

\section{Concluding remarks}

EAS simulations show that the lateral density distributions of the positive and negative muons are varying not only with the (radial) distance from the shower axis, but also with the azimuth relative to the plane of the incident shower. 
The reasons are different. In addition to the attenuation effects of charged particles of inclined showers~\cite{ref15} in the atmosphere by the variation of the traveling distances in the atmosphere, the geomagnetic field affects the travel of positive and negative muons in an opposite way. The geomagnetic effects depend on the direction of the EAS axis relative to the Earth's magnetic vector. This leads to an azimuthal variation of the muon charge ratio of the muon density distribution, which has to be regarded  in context of analyses of experimental data. In the extreme case of very inclined showers (with long slant depths) the Earth magnetic field might be used as magnetic separator~\cite{ref20,ref21}, at least for muons in the GeV range.
Obviously the experimental detection of these features is of great interest for the understanding of the EAS development. Furthermore the quantitative results would also provide some information about the hadronic interaction, in particular when observing higher energy muons. 
The aspect of the dependence on the hadronic interaction models, currently en vogue, has not been systematically explored in the present paper.

{\ack
We are indebted to R. Engel, T. Huege, D. Heck and A.W. Wolfendale for clarifying discussions about different aspects of the topic. We thank J. Oehlschl\"ager and A. Patrascioiu for substantial help in achieving the reported results. 
These studies have been prompted by the experimental plans of the Cosmic Ray Group of NIPNE, Bucharest, we are collaborating with in this field.}

\end{document}